\documentclass{osa-article}
\journal{oe}
\articletype{Research Article}

\newcommand{\p}{\partial}
\newcommand{\f}{\text{f}}
\newcommand{\h}{\text{h}}

\newcommand{\ep}{\varepsilon}
\newcommand{\vta}{\vartheta}
\newcommand{\om}{\omega}

\newcommand{\ta}{\theta}

\newcommand{\wh}{\widehat}

\newcommand{\cP}{{\cal P}}
\newcommand{\cH}{{\cal H}}
\newcommand{\cF}{{\cal F}}

\newcommand{\cW}{{\cal W}}

\newcommand{\be}{\begin{equation}}                                       
	\newcommand{\ee}{\end{equation}}
\newcommand{\ba}{\begin{eqnarray}}
	\newcommand{\ea}{\end{eqnarray}}
\newcommand{\bref}[1]{(\ref{#1})}

\newcommand{\bi}[1]{\bibitem{#1}}\newcommand{\lab}[1]{\label{#1}}

\newcommand{\bsub}{\begin{subequations}}                      
		\newcommand{\esub}{\end{subequations}}     

\begin{document}

\title{Sech-squared Pockels solitons in the  microresonator parametric down-conversion}
\author{Dmitry V. Skryabin \authormark{1,2,*}}
\address{
\authormark{1}Department of Physics, University of Bath, Bath BA2 7AY, UK \\
\authormark{2}Russian Quantum Center, Skolkovo 121205, Russia\\
\authormark{*}\textcolor{blue}{d.v.skryabin@bath.ac.uk}
}

%
\begin{abstract}
We present an explicit sech-squared-soliton solution associated with the optical Pockels effect, achieved through the generation of the frequency combs via parametric down-conversion in  optical microresonators with quadratic nonlinearity. This soliton contrasts the parametric sech-soliton describing the half-harmonic field  in the limit of the large index mismatch, and associated with the  cascaded-Kerr effect. We predict differences in the spectral profiles and powers of the Pockels and cascaded-Kerr solitons, and report that the pump power threshold of the former agree with the recent experimental observations.
\end{abstract}

	\tableofcontents

\section{Introduction}
Nonlinear and quantum optics in ring microresonators have been attracting a great deal of attention over the past decades~\cite{rev0}.  In particular, the frequency comb generation and soliton photonics in  microresonators are reaching unprecedented levels of practical relevance~\cite{rev3}. 
The most ubiquitous microresonator solitons remain  the dissipative version of the Kerr solitons, which, thanks to the small losses of modern microresonators, remain well approximated by the fundamental bright solitons of the Nonlinear Schr\"odinger (NLS) equation~\cite{rev3,mil1,npkip}. 

The second-order, $\chi^{(2)}$, nonlinearity has always been a viable alternative to the Kerr one. 
The $\chi^{(2)}$ microresonators have the potential to extend comb generation to  new wavelengths, elevate dispersion constraints, and reduce power requirements~\cite{conf}.
Using of the  $\chi^{(2)}$ response comes, however,  with the caveats of 
the need to care about the phase and group velocity matching to take  full advantage of it.   
One of the recent highlights in this area was the result by Bruch and colleagues~\cite{bru} demonstrating 
the solitons due to parametric down-conversion  in an integrated $\chi^{(2)}$ microresonator.  
A sufficiently comprehensive list of references on the solitons in $\chi^{(2)}$  
resonators can be found in~\cite{josab}. 
For further recent and historic theoretical contributions on the bright and dark 
solitons due to parametric-down conversion  
see, e.g., Refs.~\cite{ol,wlo,mash,pra2,pra1,she,longhi,stal,pre0,preold}.

While Ref.~\cite{bru} has used, or perhaps even coined, the term 'Pockels solitons', 
it is also known that the  impact of the $\chi^{(2)}$ nonlinearity on the refractive 
index change depends on how well the index matching between the pump and either half- or second-harmonic is arranged. In the  mismatched limit, the $\chi^{(2)}$ susceptibility starts mimicking the Kerr effect, which could be called the cascaded-Kerr nonlinearity, see, e.g.,~\cite{josab}. Thus, the transition between the 
Pockels and the cascaded-Kerr solitons is a problem that requires further considerations. 
Our focus here is on the bright soliton pulses.
It is known that, in the phase-mismatched cascaded-Kerr limit, the $\chi^{(2)}$ microresonator model can be reduced to the parametric NLS equation~\cite{ol,wlo}, which has the sech-soliton solution~\cite{ol,wlo,miles,fau,barash,longhi}. One outstanding problem would be finding the
 solution for the bright parametric solitons in the practically important case of the phase-matched resonators.

Any intense intra-resonator waveform produces the nonlinear shift of the resonance frequencies. For the Kerr or Kerr-like solitons, this shift is naturally expected to be proportional to the power.  
This work aims to demonstrate that, in the phase-matched microresonator parametric-down conversion, there exists an explicit bright soliton solution producing the nonlinear resonance shift proportional to the field amplitude, i.e., like should be expected in the optical Pockels effect.
The spatial profile of this soliton is given by the sech-squared function, unlike the sech-profile of the Kerr soliton. 
We note before proceeding, that the  optical Pockels effect discussed below should be distinguished from  the voltage controlled one, which is broadly used in Pockels cells, and as a frequency comb generation tool~\cite{nature,opn}.

\section{Model}
\lab{mod}
Our model of the multi-mode half-harmonic generation in a microresonator 
follows the approach of Ref.~\cite{josab}. The fundamental (pump)
field, 'f',  is spectrally centred around the pump laser frequency, 
$\om_p$,  and the half-harmonic field, 'h', is around $\om_p/2$. The fields and 
detunings are defined as
\be
\begin{split}
	&\text{fundamental/pump:}~~~~ 
	\psi_\f(\vta,t)\times \exp\left\{i2\vta J-it\om_p \right\}+c.c.,
	\\
	&\text{half-harmonic:}~~~~~~~~~~~\psi_\h(\vta,t)\times 
	\exp\left\{i\vta J-\frac{it\om_p}{2} \right\}+c.c.,
	\\
	&\text{half-harmonic detuning:}~~\om_{0\h}-\frac{\om_p}{2}=\delta,
	\\
	&\text{pump detuning:}~~~~~~~~~~~~~~~~\om_{0\f}-\om_p=2\delta-\ep.
\end{split}
\lab{opo}
\ee
Here, $\psi_\f$ and $\psi_\h$ are the envelope functions, 
$\vta\in[0,2\pi)$ is the angular coordinate along the ring, and
$J$  and $2J$ are the resonator mode numbers with the frequencies $\om_{0\h}$ and $\om_{0\f}$,
respectively.
$\ep$~is the frequency mismatch parameter,
\be
\ep=2\om_{0\h}-\om_{0\f}=
2J \frac{c}{R}
\left[\frac{1}{n_{J}}-\frac{1}{n_{2J}}\right].
\lab{en22}
\ee
Here, $n_{J}$ is the effective refractive index felt by the mode $J$, 
$c$ is the vacuum speed of light, and $R$ is the resonator radius.
Requiring $\ep=0$, 
yields the anticipated  index matching condition, $n_{J}=n_{2J}$. 
Refractive index, and hence $\ep$, can be fine-tuned by, e.g., temperature 
or electro-optic controls. Small nonlinear shifts of the refractive index proportional to either power
(optical Kerr effect) or  the field amplitude (optical Pockels effect) 
translate, via the standard Taylor series, to the respective shifts of the resonance frequencies.

Dispersion engineering is important for any type of modelocking, 
including the soliton modelocking. We set 
$D_{1\f, 1\h}/2\pi$ as the repetition-rate (free-spectral-ranges, FSR),  
and $D_{2\f, 2\h}$ as the dispersion parameters having units of Hz~\cite{josab}.
The FSR difference, $D_{1\f}-D_{1\h}$, scales inversely with $R$, while 
$D_{2\f, 2\h}\sim 1/R^2$~\cite{mil}. 
Therefore going from the mm to 10's of micron  radii leads to the relative reduction of the impact of the FSR difference on the dynamics, $|D_{1\f}-D_{1\h}|/|D_{2\f}|\sim R$. 
The group-velocity matching can be implemented either across the zero-dispersion point for the modes of the same family or, for the same signs of $D_{2\f,2\h}$, using the different mode families and the avoided mode-crossings, see, e.g.,~\cite{mode2,mode1,mode5,modes}.   Applications of the quasi-phase matching~\cite{mode6,mode7} and  optical fibres~\cite{mode3,mode4,mar}
offer further avenues for dispersion engineering in $\chi^{(2)}$ resonators. 
In order to derive transparent analytical results for Pockels solitons, we consider the case of the FSR, i.e., group velocity, matching, $D_{1\f}=D_{1\h}=D_1$,  for the same dispersion signs, i.e., $D_{2\f}/D_{2\h}>0$. The case of the opposite dispersion signs, $D_{2\f}/D_{2\h}<0$, is also proceedable analytically, but for the reasons of brevity we prefer to keep this work focused on the former. 

Transforming to the rotating frame of reference,
$\ta=\vta-D_1 t$, the equations for $\psi_{\f,\h}$ are~\cite{josab}
\bsub
\lab{mm}
\begin{align}
	i\p_t \psi_\h &=\delta \psi_\h-\tfrac{1}{2}D_{2\h}\p^2_\ta \psi_\h
	-\gamma_{\h} \psi_\f \psi_\h^* -i\tfrac{1}{2}\kappa_\h\psi_\h,
	\lab{mma}
	\\ 
	i\p_t \psi_\f&=(2\delta-\ep)\psi_\f-
	\tfrac{1}{2}D_{2\f}\p^2_\ta \psi_\f -\gamma_{\f} \psi_\h^2 -
	i\tfrac{1}{2}\kappa_\f\big(\psi_\f-\cH\big),\lab{mmb}
\end{align} 
\esub
where  $\kappa_{\f,\h}$ are the linewidths, and
$\cH$ is the pump parameter  defined as
\be
\cH=\sqrt{\frac{\eta\cF\cW}{\pi}}=\frac{1}{\kappa_\f}\sqrt{\frac{\wh\kappa_\f D_{1\f}\cW}{\pi}}. 
\lab{pump}
\ee
Here, $\eta=\wh\kappa_\f/\kappa_\f<1$ is the coupling coefficient, $\wh\kappa_\f$~is the intrinsic resonator linewidth,  and $\cF=D_{1\f}/\kappa_\f$ is finesse. $\cW$ is the 'on-chip' laser power. 
$|\psi_{\f, \h}|^2$  have units of Watts, and the  
nonlinear parameters $\gamma_{\f, \h}/2\pi\sim\chi^{(2)}$ 
are measured in Hz$/\sqrt{\text{W}}$~\cite{josab}.
Note, that $\kappa_\f\cH$ does not depend on~$\kappa_\f$.

While the results derived below could be a guideline for a range of devices, our  
choice of parameters for the numerical estimates 
is geared towards the integrated resonators as used in Ref.~\cite{bru}. 
Namely, we assume $\kappa_{\f,\h}/2\pi=100$MHz,  $D_{1\f,1\h}/2\pi=300$GHz, 
$D_{2\f,2\h}/2\pi=30$MHz, $\gamma_{\f,\h}/2\pi=500$MHz/$\sqrt{\text{W}}$, and $\eta=1/2$.
The linewidth here is two orders of magnitudes larger than in the high-Q bulk resonator samples~\cite{rev0},
this explains relatively high pump power, $\cW=80$mW, required to generate solitons in Ref.~\cite{bru}.
The sech-squared soliton derived below fits well into the above range of the parameter values,
while the sech one requires even higher input powers $\cW\sim 1$W. This is because the Pockels nonlinear response is achieved under the phase matching conditions, i.e., $|\ep|/\kappa_\h\sim 1$, while the sech-soliton is associated with the cascaded-Kerr nonlinear response triggered for $|\ep|/\kappa_\h\sim 10^2$. The large $|\ep|$ lead to the inefficient conversion and  push the soliton thresholds up.

\section{Sech-squared soliton: Pockels regime}
\lab{sech2}
To find the Pockels soliton we first neglect the loss terms by assuming $|\delta|\gg\kappa_\h$, $|2\delta-\ep|\gg\kappa_\f$. Practically, it suffices to take $|\delta|/\kappa_\h\sim 10$ and 
$|\ep|/\kappa_\h\lesssim 1$.  Then, Eq.~\bref{mm} become 
\bsub
\lab{sm2}
\begin{align}
	\delta \psi_\h & -\tfrac{1}{2}D_{2\h}\p^2_\ta \psi_\h
	-\gamma_{\h} \psi_\f \psi_\h^* =0,
	\lab{sm2a}
	\\ 
	(2\delta-\ep)\psi_\f & -
	\tfrac{1}{2}D_{2\f}\p^2_\ta \psi_\f -\gamma_{\f} \psi_\h^2 =
	i\tfrac{1}{2}\kappa_\f\cH.\lab{sm2b}
\end{align} 
\esub

The soliton solution is sought in the form
\bsub
\lab{ex0}
\begin{align}
&	\psi_\h=B~\psi(\ta) e^{i\phi/2},
\lab{ex0a}\\ 
& \psi_\f=\psi(\ta) e^{i\phi}-i H,
\lab{ex0b}
\end{align}
\esub
so that the half-harmonic pulse has the vanishing background and the fundamental has the finite one.
Here, the soliton profile $\psi(\ta)$,  the phase $\phi$, the dimensionless constant $B$, and 
the  background field $H$, should be determined. We note that the existence of the zero solution (vanishing background) for $\psi_\h$ also holds for the full model, see Eq.~\bref{mm}.

We substitute Eq.~\bref{ex0} in Eqs.~\bref{sm2}, and
separate the real and imaginary parts. 
The first outcomes of this procedure are the explicit expressions for the soliton background
\be
	  H =\frac{\kappa_\f\cH }{2(2\delta-\ep)},\lab{x1}
	\ee  
and for the phase, $\phi=\pm\pi/2$. The $\phi=-\pi/2$ makes the soliton part of $\psi_\f$ in Eq.~\bref{ex0b} to be in-phase with the background field, which is known to correspond to the always unstable soliton family. The $\phi=\pi/2$ solution is $\pi$-out-of-phase with the background and is a largely stable one~\cite{pre0,fau}. 

The further result of the substitution  is a pair of the differential equations for  $\psi$,
\bsub
\lab{a3}
\begin{align}
\delta\psi-&\tfrac{1}{2}D_{2\h}
\p^2_\ta \psi	-\gamma_{\h}\psi^2=0,\lab{a3b}\\
(2\delta-\ep)\psi-	& \tfrac{1}{2}D_{2\f}\p^2_\ta\psi -
\gamma_{\f}B^2 \psi^2 =0.
	\lab{a3a}
\end{align}
\esub
The solution of Eq.~\bref{a3b} is defined here as the sech-squared Pockels soliton,
\be
\psi=\frac{3\delta}{2\gamma_\h}\text{sech}^2\left(\ta\sqrt{\frac{\delta}{2D_{2 \h}}}\right),
~\delta D_{2 \h}>0.
\lab{ex1b}
\ee

The validity of Eq.~\bref{ex1b} can be verified by substitution.
Thus, if dispersion is normal, then the sech-squared solitons exist for 
$\delta<0$ and,  for $\delta>0$, if dispersion is anomalous. 
The sech-squared soliton can be found among the in-line equations in the
paper by Karamzin and Sukhorukov on the beam diffraction in a 
$\chi^{(2)}$ crystal~\cite{kara}, and it also 
has been later found and elaborated on by others, see, e.g., \cite{dru,pr}. The results derived by us here 
are a generalization of Ref.~\cite{kara} that incorporates the pump term  (this section) and the linewidth (section~\ref{limit}), and thereby shows how the sech-squared solitons become relevant in the resonators, in general, and microresonators, in particular.

$\psi$ in Eq.~\bref{ex1b} must simultaneously solve Eqs.~\bref{a3a} and \bref{a3b}, which 
requires acknowledging the following conditions
\be
\frac{\delta}{D_{2\h}}=\frac{2\delta-\ep}{D_{2\f}},~
	\frac{\gamma_\h}{D_{2\h}}=\frac{\gamma_{\f}B^2}{D_{2\f}}.\lab{bb}
\ee
The latter of the above fixes the, so far free, parameter $B$, see Eq.~\bref{ex0}, 
and  implies that the dispersions 
are either both normal or both anomalous, i.e., $D_{2\h}/D_{2\f}>0$.  
The first of the conditions in Eq.~\bref{bb} restricts the combination of the detuning, frequency mismatch, and dispersions necessary for the analytic solution to exist. 

We first look into how the peak power of the sech-squared soliton scales with the detuning by taking  
the important practical case of the exact phase matching, $\ep=0$.
Fixing dispersions to comply with Eq.~\bref{bb},  $D_{2\f}/D_{2\h}=2$, 
is a soft assumption to make for the sake of dealing with a transparent analytical solution.
For  $\delta/\kappa_\h =10$,  the peak soliton power 
in the fundamental and half-harmonic fields are
\bsub
\lab{e5}
\begin{align}
	&
	\max|\psi_\f|^2=\frac{9\delta^2}{4\gamma_\h^2}\approx 9\text{W},
	\lab{e5a}\\
	&
	\max|\psi_\h|^2=\frac{9\delta^2}{2\gamma_\h^2}\approx 18\text{W}.
	\lab{e5b}
\end{align}
\esub
For $\cW= 80$mW, we find that $\cH^2\approx 40$W, and    
the power of the soliton background is 
$\kappa_\f^2\cH^2/16\delta^2=25$mW, see Eq.~\bref{x1}.  
We note that the peak powers 
of the half-harmonic pulse are proportional to $\delta^2$, hence,  $ \max|\psi_\h|\sim \delta=\om_{0\h}-\om_p/2$, and therefore the  shift $\om_{0\h}$ and the effective index change, see Eq.~\bref{en22}, induced by the soliton are proportional to the field amplitude, so that, we are indeed dealing with the Pockels regime of the resonator operation.
 
  \begin{figure}[t]
 	\centering{	
 		\includegraphics[width=0.45\textwidth]{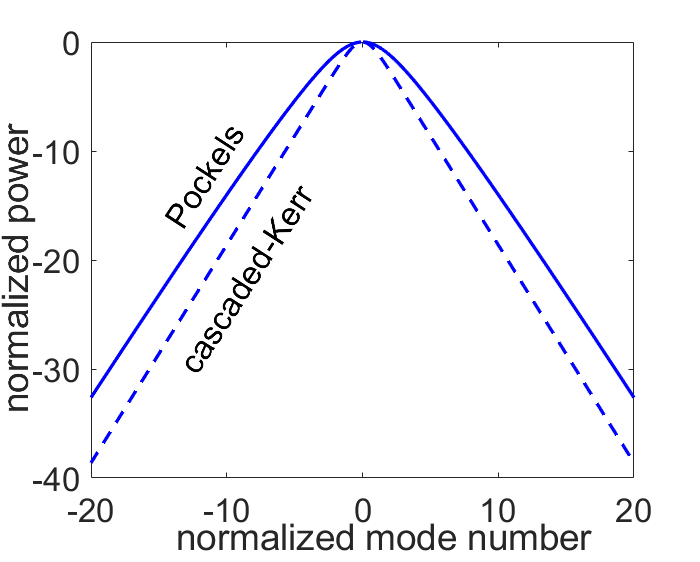}	
 	}
 	\caption{The normalized power spectra for the Pockels (sech-squared) and cascaded-Kerr (sech) soliton solutions vs the normalised mode number, $m=\mu/\mu_0$. The full line is $\text{log}\left(m^2/\sinh^2m\right)$, see Eq.~\bref{sp1}, and the dashed one is 
 		$\text{log}\left(\text{sech}^2m\right)$, see Eq.~\bref{sppp1}.	}
 	\label{f1}
 \end{figure}

The total power and the spectrum of the half-harmonic component of the sech-squared Pockels soliton are
\bsub
\begin{align}
\cP= B^2\int^{2\pi}_{0} |\psi_\h|^2 d\ta&=
\cP_\text{Pock}~\left[\frac{|\delta|}{\kappa_\h}\right]^{3/2},
\lab{pk1}\\
\cP_\text{Pock}&\approx\frac{6\kappa_\h\sqrt{2\kappa_\h D_{2\h}}}{\gamma_\h^2}\approx 0.2\text{W},
\lab{pk1b}
\end{align}
\esub
and
\be
S_{\mu}\sim\Big\vert\int_{0}^{2\pi}
\psi_\h e^{i\mu\ta} \frac{d\ta}{2\pi}\Big\vert^2
\sim \left(\frac{\mu/\mu_0}{\text{sinh}~\mu/\mu_0}\right)^2,~~~\mu_0=\frac{1}{\pi}
\sqrt{\frac{\delta}{2D_{2 \h}}},
\lab{sp1}
\ee
respectively. The  integrals above have been calculated by extending the zero-$2\pi$ interval 
to the infinity  and   by applying the table integrals found in Ref.~\cite{ryzh}. The difference of $S_\mu$ in Eq.~\bref{sp1} with 
the triangular spectrum of the sech-soliton is illustrated in Fig.~\ref{f1}, and 
the comparative discussion is included in Section~\ref{sech}.

\section{Limits of existence of sech-squared solitons}
\lab{limit}
Accepting a small level of complication of the algebra as a necessity lets us to trace the effects of the so far neglected loss terms. This needs to 
be limited  by  the half-harmonic loss, $\kappa_\h$,
which, however,  reveals the net effect well, and allows 
to remain within the comfortably transparent analytical considerations. 
Eqs.~\bref{sm2} are now replaced with
\bsub
\lab{xsm2}
\begin{align}
\delta \psi_\h&-\tfrac{1}{2}D_{2\h}\p^2_\ta \psi_\h
-\gamma_{\h} \psi_\f \psi_\h^* =i\tfrac{1}{2}\kappa_\h\psi_\h,
\lab{xsm2a}
\\ 
(2\delta-\ep)\psi_\f&-
\tfrac{1}{2}D_{2\f}\p^2_\ta \psi_\f -\gamma_{\f} \psi_\h^2 =
i\tfrac{1}{2}\kappa_\f\cH.\lab{xsm2b}
\end{align} 
\esub
The substitution is again as in Eq.~\bref{ex0} and the background amplitude as in Eq.~\bref{x1}.  
The equations for the phase become nontrivial
\begin{equation}
\cos\phi=\frac{\kappa_\h(2\delta-\ep)}{\kappa_\f\gamma_\h \cH},~
\sin\phi=\pm\sqrt{1-\cos^2\phi}
\lab{xx2}
\end{equation}
In the limit $\kappa_\h\to 0$ considered in the previous section, the $\sin\phi=-\sqrt{~}$ state corresponds to $\phi\to-\pi/2$, i.e.,  the soliton is in-phase with the background field, which is 
the always unstable soliton family. The $\sin\phi=+\sqrt{~}$, $\phi\to \pi/2$ soliton 
is $\pi$-out-of-phase with the background and can be stable~\cite{pre0,fau},
see similar comments after Eq.~\bref{x1}. 

Eqs.~\bref{ex0}, \bref{xsm2}, \bref{xx2} lead 
to the refreshed system for  $\psi$ 
\bsub
\lab{xa3}
\begin{align}
	(2\delta-\ep)\psi & -	\tfrac{1}{2}D_{2\f}\p^2_\ta\psi -
	\gamma_{\f}B^2 \psi^2 =0,
	\lab{xa3a}\\
	\left(\delta+\tfrac{1}{2}\kappa_\h\tan \phi\right)\psi & -\tfrac{1}{2}D_{2\h}
	\p^2_\ta \psi	-\gamma_{\h}\psi^2=0.
	\lab{xa3b}
\end{align}
\esub

 \begin{figure}[t]
	\centering{	
		\includegraphics[width=0.45\textwidth]{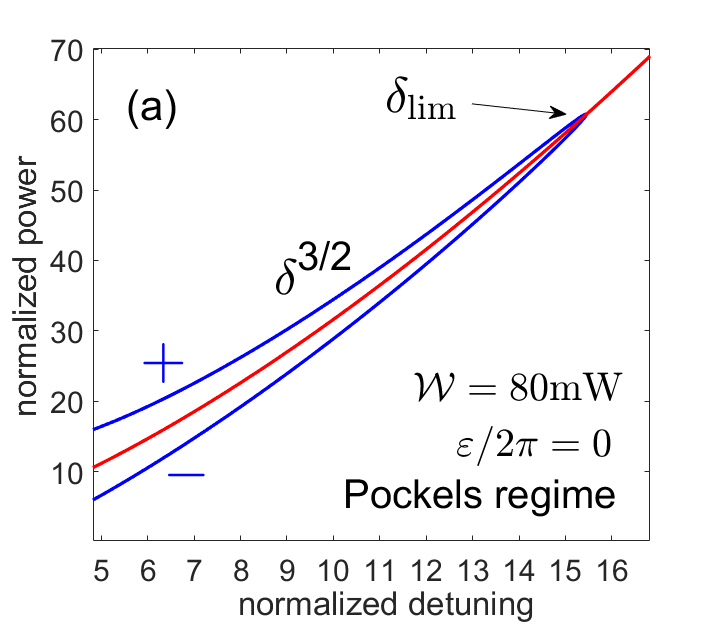}		
		\includegraphics[width=0.45\textwidth]{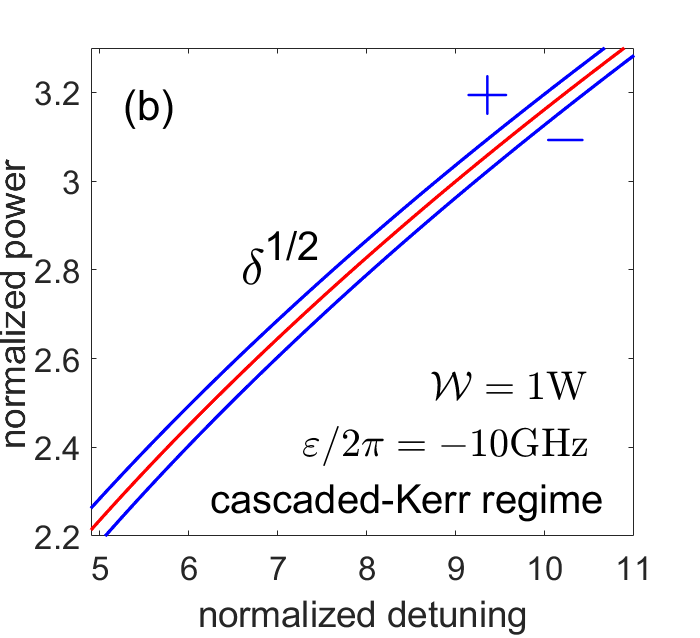}	
	}
	\caption{The normalised total power of the half-harmonic soliton vs the normalized detuning, $\delta/\kappa_\h$.
		(a)~is the sech-squared soliton case. The red line shows $\cP/\cP_\text{Pock}=|\delta/\kappa_\h|^{3/2}$,  $\cP_\text{Pock}\approx 0.2$W, $\ep/2\pi=0$. The blue lines show the effect of the losses, see Section~\ref{limit}.
		The plus and minus branches correspond to the respective signs in Eq.~\bref{xex1b}. (b)~is the sech soliton case. The red line shows $\cP/\cP_\text{cKerr}=|\delta/\kappa_\h|^{1/2}$, $\cP_\text{cKerr}\approx 6$W, $\ep/2\pi=-10$GHz. The plus and minus branches correspond to the respective signs in the in-line $\bar\delta$ equation in the first sentence of Section 6.
	}
	\label{f2}
\end{figure}

The soliton solution of Eq.~\bref{xa3b} is then,
\begin{equation}
\psi=\frac{3\bar\delta}{2\gamma_\h}
\text{sech}^2\left(\ta\sqrt{\frac{\bar\delta}{2D_{2 \h}}}\right),~~ 
\bar\delta=\delta+\frac{\kappa_\h}{2}\tan \phi,
\lab{xex1a}
\end{equation}
where $\bar\delta$ takes two values, see Eq.~\bref{xx2},
\begin{equation}
\bar\delta_\pm=\delta\pm\frac{\kappa_\h}{2}
\sqrt{\frac{\kappa_\f^2\gamma_\h^2\cH^2}{\kappa_\h^2(2\delta-\ep)^2}-1}.
\lab{xex1b}
\end{equation}
$\bar\delta_+=\bar\delta_-=\delta_{\lim}$ when  the square root becomes zero, 
\be
\delta_{\lim}=\pm\frac{\kappa_\f\gamma_\h\cH}{2\kappa_\h}+\frac{\ep}{2},
\ee
which sets the existence limit for the sech-squared solitons.
If, e.g.,  $D_{2\f,2\h}>0$ (anomalous dispersion) and $\ep= 0$, 
then one should choose plus in the above. The compatibility of  Eqs.~\bref{xa3a} and ~\bref{xa3b},
requires replacing the first condition in Eq.~\bref{bb} with 
$\bar\delta/D_{2\h}=(2\delta-\ep)/D_{2\f}$. 
Plots of the Pockels soliton powers, that account (blue) and disregard (red) $\kappa_\h$, are shown in Fig.~\ref{f2}(a).
The input power is chosen to be $80$mW as in Ref.~\cite{bru}, which also 
provides the soliton existence around $\delta/\kappa_\h\approx 10$ fitting with our assumptions.

\section{Sech soliton: cascaded-Kerr regime}
\lab{sech}
The parametric sech-solitons in resonators~\cite{longhi,ol} are like the ones previously found  in
a few other physical contexts, see, e.g., Refs.~\cite{miles,fau,barash,longhi}. They 
can be derived via the reduction of Eq.~\bref{mm} to the parametric NLS equation. 
The reduction starts from the assumption that the resonator is pumped well away from the resonance, $|\ep|\gg|\delta|$, while the half-harmonic is generated close to the resonance. 
This practically inefficient phase-mismatched arrangement allows to approximately resolve Eq.~\bref{mmb} with 
\be
\psi_\f\approx\frac{\gamma_\f\psi_\h^2-i\tfrac{1}{2}\kappa_\f \cH}{-\ep},
\lab{e1}
\ee 
so that 
Eq.~\bref{mma} becomes
\be
i\p_t \psi_\h=\delta \psi_\h-\frac{1}{2}D_{2\h}\p^2_\ta \psi_\h 
+  \frac{\gamma_\f\gamma_{\h}}{\ep}|\psi_\h|^2\psi_\h
-  i\frac{\kappa_\f\gamma_\h \cH}{2\ep} \psi_\h^*-i\frac{1}{2}\kappa_\h\psi_\h.
\lab{pe}
\ee
Now, the half-harmonic  experiences the effective Kerr, i.e.,
cascaded-Kerr, nonlinearity. For $\ep<0$,
the cascaded-Kerr effect is positive, i.e., focusing, and for $\ep>0$ it is negative. 
To be consistent with Section~\ref{sech2}, we first  neglect the $\kappa_{\f,\h}$ terms, which works well if
$|\ep|\gg|\delta|\gg\kappa_{\f,\h}$, e.g., for $\ep/\kappa_\h=-10^2$ and $\delta/\kappa_\h =10$.
Then, the exact soliton solution of  Eq.~\bref{pe} is
\be
\psi_\h=\sqrt{\frac{-2\delta\ep}{\gamma_\f\gamma_\h}}\times\text{sech}\left(\ta\sqrt{\frac{2\delta}{D_{2\h}}}\right)
e^{\pm i\pi/4},~\delta\ep<0, ~\delta D_{2\h}>0.
\lab{e2}
\ee

For the sech solitons to exist, the signs of dispersion, detuning, and of the mismatch parameters must be correlated as indicated above. The peak  powers of the cascaded-Kerr soliton are 
\bsub
\lab{e4}
\begin{align}
	&\max|\psi_\f|^2=\frac{4\delta^2}{\gamma_\h^2}
	\approx 16\text{W},
	\lab{e4a}\\
	&\max|\psi_\h|^2=
	\frac{2|\delta||\ep|}{\gamma_\f\gamma_\h}\approx 80\text{W}.
	\lab{e4b}
\end{align}
\esub
Now, the peak power of the half-harmonic, $\max|\psi_\h|^2$, is growing proportionally to $\delta$, cf., Eq.~\bref{e5}. This is the same as for the non-parametric NLS solitons in Kerr microresonators~\cite{rev3,npkip},
and is consistent with the change of the effective  index, $n_J$, being proportional
to the pump power as it should be expected in the optical Kerr effect. The intrinsic $\chi^{(3)}$ effect typically becomes comparable with the cascaded nonlinearity if the mismatch parameter becomes orders of several FSRs~\cite{josab}, while here we keep $\ep/2\pi$ to be much less than FSR. For studies of the Kerr 
soliton combs supported by the intrinsic $\chi^{(3)}$ nonlinearity while  $\chi^{(2)}\ne 0$ see, e.g., \cite{micro3,pow,gaeta,nie}.

The total power and spectrum for the   cascaded-Kerr solitons are
\bsub
\begin{align}
	\cP= \int^{2\pi}_{0} |\psi_\h|^2 d\ta&=
	\cP_{\text{cKerr}}~\left[\frac{ |\delta|}{\kappa_\h}\right]^{1/2},
	\lab{k1}\\
	\cP_\text{cKerr}&\approx\frac{|\ep|\sqrt{8\kappa_\h D_{2\h}}}{\gamma_\h\gamma_\f}\approx 6\text{W},
	\lab{k1b}
\end{align}
\esub
and 
\be
S_{\mu}\sim
\left(\text{sech}\frac{\mu}{\mu_0}\right)^2,
~\mu_0=\frac{1}{\pi}
\sqrt{\frac{2\delta}{D_{2 \h}}}.
\lab{sppp1}
\ee
The logarithm of $S_{\mu}$ in Eq.~\bref{sppp1} has the triangular shape, while
in the index-matched case (Pockels limit), the top of the spectrum  is more rounded
with the tails carrying more power, 
which is a reflection of the $\mu^2$ factor in Eq.~\bref{sp1}, see  Fig.~\ref{f1}. 

\section{Pump power threshold: Sech-squared vs sech solitons}
Retaining the $\kappa_\h$-loss term in Eq.~\bref{pe}, 
leads to the substitution $\delta\to\bar\delta=\delta\pm\sqrt{\kappa_\f^2\gamma_\h^2\cH^2/\kappa_\h^2\ep^2-1}$ 
in Eq.~\bref{e2}. 
Plots of the soliton power, that account (blue) and disregard (red) $\kappa_\h$, are shown in Fig.~\ref{f2}(b).
Because of the
dominance of the index-matching parameters $\ep$ in this regime,  the square-root  is independent from $\delta$. The existence of the cascaded-Kerr soliton requires  $\cH>\cH_\text{cKerr}=|\ep|\kappa_\h/\gamma_\h\kappa_\f$, i.e., the laser power is
\be
\cW>\cW_\text{cKerr}=\frac{|\ep|^2\kappa_\h^2}{\gamma_\h^2\kappa_\f^2}\frac{\pi}{\eta\cF}\approx 1\text{W},
\lab{p1}
\ee
where $\ep/2\pi=-10$GHz, i.e., $\ep/\kappa_\h=-100$.
For $|\ep|$ approaching $100$GHz,  the more careful sech-soliton threshold power estimates would require accounting for the intrinsic Kerr effect, see, e.g.,~\cite{bru,josab,ol,micro3,pow,gaeta,nie}. This goes beyond the main objective of the present communication, which is the theory of the sech-squared (Pockels) solitons existing for $\ep\sim\kappa_\h$.

If $\ep=0$, or $|\ep|\ll|\delta|$, then  the Pockels soliton given by Eqs.~\bref{xex1a}, \bref{xex1b} exists for the much smaller  threshold powers determined by the detuning, 
\be
\cW>\cW_\text{Pock}=\frac{4|\delta|^2\kappa_\h^2}{\gamma_\h^2\kappa_\f^2}\frac{\pi}{\eta\cF}\approx 30\text{mW},
\lab{p2}
\ee
where $\delta/\kappa_\h=10$.
We recall that Ref.~\cite{bru} has used $80$mW of the laser power.

\section{Summary}

We have elaborated the explicit sech-squared Pockels solitons describing the soliton regime of the phase and group velocity matched parametric down-conversion in microresonators with either normal or anomalous dispersion. We have compared these solitons with the   cascaded-Kerr sech-solitons existing away from the phase-matching condition and revealed the 
differences  in (i)~the values of the threshold powers, see Eqs.~\bref{p1}, \bref{p2}, 
(ii)~scaling of the soliton power with the detuning parameter, see Fig.~\ref{f2}, and 
(iii)~how the Pockels-soliton spectra  deviate from the triangular spectra of the sech-solitons, see Fig.~\ref{f1} and, cf.,~Eqs.~\bref{sp1} and \bref{sppp1}. 
The sech and sech-squared shapes of the half-harmonic field are 
embedded, as the two limit cases, inside the wider and numerically accessible family of the bright soliton solutions, see, e.g.,~\cite{bru,ol,wlo,stal,pre0,preold}. 

\section*{Funding}  
Russian Science Foundation (17-12-01413-$\Pi$).

\section*{Disclosures}
The author declares no conflicts of interest.

\section*{Data availability} 
No data files are associated with this manuscript.

\end{document}